\documentclass[a4paper]{article}
\usepackage{amsmath}
\usepackage[english]{babel}
\usepackage{latexsym}
\usepackage{amssymb}
\usepackage{amscd}
\usepackage{amsgen,amstext,amsbsy,amsopn}
\usepackage{math rsfs}

\usepackage{amsthm,epsfig,graphicx,graphics}
\usepackage[latin1]{inputenc}
\usepackage{xspace}
\usepackage{amsxtra}

\newcommand{\half}{\mbox{$\frac{1}{2}$}}
\newcommand{\x}{\mathbf {x}}
\newcommand{\xv}{\mathbf {x}}

\newcommand{\lf}{\left}
\newcommand{\ri}{\right}
\newcommand{\eps}{\varepsilon}
\newtheorem{teo}{Theorem}

\newcommand{\version}{December 15, 2012}
\begin{document}
\title{Vortex Phases of Rotating Superfluids\footnote{For the Proceedings of 21st INTERNATIONAL LASER PHYSICS WORKSHOP, Calgary, July 23-27, 2012}}

\author{M. Correggi	\\ \normalsize\it Dipartimento di Matematica,
Universit\`{a} degli Studi Roma Tre,\\ \normalsize\it
Largo San Leonardo Murialdo 1, 00146, Roma, Italy. \\ \normalsize\it \hspace{-.5 cm}	\\ F. Pinsker \\	\normalsize\it DAMTP, University of Cambridge, Wilbertforce Road, Cambridge CB3 0WA, UK\\ \normalsize\it \hspace{-.5 cm} \\	N. Rougerie	\\	\normalsize\it	Universit\'e Grenoble 1 \& CNRS,  LPMMC (UMR 5493), B.P. 166, 38 042 Grenoble, France
 \\ \normalsize\it \hspace{-.5 cm} \\   J. Yngvason	\\ \normalsize\it Erwin Schr{\"o}dinger Institute for Mathematical Physics,	Boltzmanngasse 9, 1090 Vienna, Austria,	\\ \normalsize\it Fakult\"at f\"ur Physik, Universit{\"a}t Wien,	 Boltzmanngasse 5, 1090 Vienna, Austria.}
 \date{\version}
\maketitle

\begin{abstract} We report on the first mathematically rigorous proofs of a transition to a giant vortex state of a superfluid in rotating  anharmonic traps.  The analysis is carried out within two-dimensional Gross-Pitaevskii theory at large coupling constant and large rotational velocity and is based on precise asymptotic estimates on the ground state 
energy. An interesting aspect is a significant difference between \lq soft' anharmonic traps (like a quartic plus quadratic trapping potential) and traps with a fixed boundary. In the former case vortices persist in the bulk until the width of the annulus becomes comparable to the size of the vortex cores. In the second case
the transition already takes place in a parameter regime where the size of vortices is very small relative to the width of the annulus. Moreover, the density profiles in the annulus are different in the two cases. In both cases rotational symmetry of the density in a true ground state is broken, even though a symmetric variational ansatz gives an excellent approximation to the energy.

\end{abstract}

\section{Introduction}

A superfluid confined in a rotating anharmonic trap, where the rotation speed can in principle be arbitrarily large,  undergoes several phase transitions as the speed increases.  At first the fluid is vortex free \cite{IM1,AJR} but then quantized vortices emerge, eventually forming a vortex lattice \cite{CD,  Fe1, AD, IM2,  A, Co, Fe2} that may persist even when the speed is so large that the centrifugal force creates a `hole' with strongly depleted density in the middle of the trap \cite{FB, CDY1, CDY2, CY}. Above a certain rotation speed a transition to a {\it giant vortex state} takes place. In this state the vortices disappear from the annulus where the bulk of the superfluid is concentrated while  a macroscopic phase circulation remains. This phenomenon has been studied theoretically by variational and numerical methods in the past \cite{Fe2, FJS,FB,FZ,KTU,KF, DK} but mathematically rigorous proofs of the giant vortex transition have been obtained only very recently \cite{R1, CRY, CPRY1, CPRY2, CPRY3, R2}. An experimental realization of this transition appears to be still out of reach although anharmonic traps have been available already for some time \cite{BSSD,Ry,H,Sh}. In the following we report on the main findings of this analysis with emphasis on \cite{CPRY2, CPRY3}.

\section{Setting the Stage}

\subsection{The basic many-body Hamiltonian}

The quantum mechanical Hamiltonian for $N$ spinless
bosons with an {external potential},
$V$, and a {pair interaction potential}, $v$,  in a {rotating frame} with angular velocity ${\mathbf \Omega_{\rm rot}}$ is
{\begin{equation*}
H = 
\sum_{j=1}^{N} \left(- \half \nabla^2_j +V(\x_{j})-{\mathbf L}_j\cdot {\mathbf \Omega_{\rm rot}}\right)+
\sum_{1 \leq i < j \leq N} v(|\x_i - \x_j|).
\end{equation*}}
Here $\mathbf x_j\in \mathbb R^3$ and ${\mathbf L}_j=-\mathrm i \mathbf x_j\times\nabla_j$ is the angular momentum of the $j$th particle.
The Hamiltonian can alternatively  be written in the \lq magnetic' form 
\begin{equation}\label{ham}
H= 
\sum_{j=1}^{N} \left\{\half (\mathrm i\nabla_j+{\mathbf A}({\x}_j))^2+V(\x_{j})-\hbox{$\frac 12$}\Omega_{\rm rot}^2r_j^2\right\}+
\sum_{1 \leq i < j \leq N} v(|\x_i - \x_j|)
\end{equation}
with the vector potential $${\mathbf A}(\x)={\mathbf \Omega_{\rm rot}}\times \x=\Omega_{\rm rot} r\, \mathbf e_\vartheta$$ where $r$ denotes the distance from the rotation axis and $\mathbf e_\vartheta$ the unit vector in the angular direction. This way of writing the Hamiltonian corresponds to the splitting of the rotational effects into Coriolis and centrifugal forces.


\subsection{Harmonic vs.\ anharmonic traps}

If $V$ is a {harmonic} oscillator potential in the direction $\perp$ to $\mathbf \Omega_{\rm rot}$, i.e., 
$$V(\x)=\half\Omega_{\rm trap}r^2+V^{\parallel}(z)$$
then stability requires
$\Omega_{\rm rot}< \Omega_{\rm trap}$. {Rapid rotation} means here that
$$\Omega_{\rm rot}\to\Omega_{\rm trap}$$
from below. On the other hand, if  $V$ is {anharmonic} and increases {faster than quadratically} in the direction $\perp$ to $\Omega_{\rm rot}$, e.g. $V(\x)\sim r^s+V^{\parallel}(z)$ with $s>2$, then $\Omega_{\rm rot}$ can in principle be as large as one pleases and {rapid rotation} means simply $\Omega_{\rm rot}\to\infty$. 

These two cases are quite different both physically and mathematically. The former leads to an effective many-body Hamiltonian in the lowest Landau level of the magnetic kinetic energy term in \eqref{ham} and bosonic analogues of the Fractional Quantum Hall Effect (see \cite{WG, Vie,Co, LS}). In the case of rapid rotation in an anharmonic trap, as considered here,  it is usually sufficient to employ Gross-Pitaevskii (GP) theory for an effective description. We remark, however, that a small anharmonic term appropriately tuned can also lead to interesting modifications of the Quantum Hall states of harmonic traps  \cite{RSY}.


\subsection{The Gross-Pitaevskii limit theorem}

The following basic fact about the  many-body Hamiltonian \eqref{ham} for $N\to\infty$ {with} $Na$ {and} $\Omega_{\rm rot}$ {\it fixed}, where $a$ is the {scattering length} of the (repulsive, short range) interaction potential $v$ was proved in \cite{LieS}:

There is (possibly fractionated) {Bose-Einstein condensation} in the ground state as $N\to\infty$, and the wave function of the condensate 
(``superfluid order parameter") is a minimizer (in general not unique) of the {GP energy functional}
\begin{equation}\label{gpenfunc}
	\mathcal E^{\rm GP}_{\rm 3D}[\Psi] = 
	\int_{\mathbb R^3}  \: \lf\{ \half\lf|\lf( {\rm i} \nabla +\mathbf A\ri) \Psi \ri|^2 
	+(V- \half \Omega_{\rm rot}^2 r^2) |\Psi|^2 
	+ 2\pi Na\,|\Psi|^4\ri\}\end{equation}
	
with $\int|\Psi|^2=1$. The Gross-Pitaevskii  partial differential equation
\begin{equation*} \big\{\lf( {\rm i} \nabla +\mathbf A\ri)^2+(V- \half \Omega_{\rm rot}^2 r^2)+4\pi Na|\Psi|^2\big\}\Psi=\mu\Psi
\end{equation*} 
with the chemical potential $\mu$ is the variational equation corresponding to this minimization problem.

The rigorous proof of this theorem is far from simple, as can be seen from the fact that a Hartree variational ansatz for the Hamiltonian \eqref{ham} (that would anyhow only lead to an upper bound) is meaningless if the interaction potential has a hard core. Even for 'soft' potentials a naive computation would not lead to \eqref{gpenfunc} with the scattering length as parameter, but rather the integral $\int v$ (that gives only the lowest Born approximation to the scattering length). 
For the mathematical background of this and related results \cite{LSSY} may be consulted. A limit theorem that holds  uniformly the parameters $\Omega_{\rm rot}$ and $Na$ as $N\to\infty$, but is restricted to the leading order, was proved in  \cite{BCPY}.


\subsection{2D Gross-Pitaevskii theory in anharmonic traps}

The GP minimization problem  has {two parameters}, $\Omega_{\rm rot}$ and $Na$. We shall be concerned with phenomena that occur in  {anharmonic} traps in the {asymptotic regime} where {both} $\Omega_{\rm rot}$ and $Na$ are {{large}}.  
For convenience introduce 
$$\varepsilon\equiv (2\pi Na)^{-1/2}$$
which  is {{small}} if $Na$ is large. (In appropriate units $\eps$ is the `healing length'.)

For traps that are sufficiently elongated along the rotational axis ($z$-direction) the properties of the condensate are to a good approximation independent of $z$ and we may consider a {2D}\footnote{A 2D description is, of course,  also appropriate in thin traps where the motion along the $z$-axis is `frozen' \cite{SY}.} energy functional
 
	$$\mathcal E_{\rm 2D}^{\rm GP}[\Psi] = \int_{\mathbb R^2}  \: \lf\{ \half\lf|\lf( {\rm i} \nabla +\mathbf A \ri) \Psi \ri|^2 +(V- \half \Omega_{\rm rot}^2 r^2) |\Psi|^2 + \frac{|\Psi|^4}{\eps^2} \ri\}$$

with a trap potential of the form (for simplicity)
\begin{equation}\label{pot} V(r)=kr^s\end{equation}
with $s>2$, $k>0$. Here $\Omega_{\rm rot}$ can be {arbitrary large}.

The {limiting case} $s\to\infty$ corresponds to a { \lq flat' trap} with fixed boundary at $r=1$. The effective potential is then simply $-\half\Omega_{\rm rot}^2r^2$ and the integration is limited to the {unit disc} in $\mathbb R^2$. A word of caution: The limit $s\to\infty$ can not be interchanged with the limits $\eps\to 0$, $\Omega_{\rm rot}\to \infty$ as discussed in Section 3.6 below. \medskip




The analysis of the GP minimizer is guided by the following heuristics:

\begin{itemize}
\item A vortex, i.e.,  a zero of the wave function $\Psi(\x)=|\Psi(\x)|\exp(\mathrm i\theta (\x))$ with an accompanying nonzero winding number of the phase factor, reduces the kinetic energy because the associated current $\sim\nabla\theta(\x)$  compensates partly the velocity field generated by ${\mathbf A}(\x)={\mathbf \Omega_{\rm rot}}\times \x$. 

\item A vortex causes also a change in the density, however,  (mass is moved from the vortex core to the bulk)  and this increases the interaction energy that depends on the density at the potential location of the vortex. The energy balance decides whether or not a vortex is favorable, and if that is the case, the size of the vortex core. 

\item {A vortex is the more costly the higher the density}. At sufficiently high rotational velocities  the compression due to centrifugal forces creates a `hole' and the density in the bulk increases until, at some point,  vortices become too costly.{}
\end{itemize}


\subsection{Scaling of the energy functional}

The effective potential $(kr^s-\half \Omega_{\rm rot}^2r^2)$ has a unique minimum at $r=(\Omega_{\rm rot}^2/(sk))^{1/(s-2)}$. Taking this as a length unit we obtain the {scaled energy functional}
\begin{equation*}\label{gpscaled}\mathcal E^{\rm GP}[\psi]=\int_{\mathbb R^2}\left\{\half|(\mathrm i\nabla+\mathrm \Omega x\mathbf e_{\vartheta})\psi|^2 +\Omega^2W(x)|\psi|^2+\eps^{-2}|\psi|^4\right\}\end{equation*}
where $x=|\xv|$, $\Omega\sim\Omega_{\rm rot}^{(s+2)/(s-2)}$, and $$\label{effpot2} W(x)=\left(\hbox{$\frac1{s}$}x^s-\half x^2\right).$$

The scaled potential has a minimum at $x=1$, independent of $\Omega$.

\section{Analysis of the GP Minimizers}

\subsection{Critical velocities}

The basic facts  for  traps of the form \eqref{pot} with $2<s<\infty$ can be summarized as follows.

{As $\Omega$ increases there are {\bf three critical velocities:}}{}
\begin{itemize}
\item $\Omega_{\rm c1}'\sim |\log\eps|$ marking the appearance of the {first vortex}.\footnote{Here $\Omega'\sim \eps^{-4/(s+2)}\Omega_{\rm rot}$. This scaling is more convenient than $\Omega\sim\Omega_{\rm rot}^{(s+2)/(s-2)}$ for $\Omega\ll 1/\varepsilon$.}
\item $\Omega_{\rm c2}\sim \eps^{-1}$ marking the appearance of a {\lq hole'} due to the centrifugal forces.{}
\item $\Omega_{\rm c3}\sim \eps^{-4}$ marking the transition to a {\lq giant vortex'}.
\end{itemize}{}

For the first transition we refer to \cite{AD, IM1, A, D}. For $\Omega_{\rm c1}\ll \Omega\ll \Omega_{\rm c3}$ the ground state energy is well approximated by assuming  a triangular {vortex lattice} in the bulk.\footnote{The reason why a triangular arrangement with hexagonal unit cells is optimal amongst regular lattices can be made plausible by appealing to an electrostatic analogy and Newton's theorem  \cite{CY}: Hexagonal cells are as  close to being circular as possible and thus have smaller multipole moments and lower interaction energy than other cells.} In the limit $\eps\to 0$ the vorticity becomes uniformly distributed with density $\Omega$ \cite{CPRY3}. 
For $\Omega>\Omega_{\rm c3}$ the {bulk is free of vortices} but a {macroscopic circulation} around the origin remains \cite{CPRY2, CPRY3}.


\subsection{The vortex lattice regime }
The ground state energy for $\Omega_{\rm c1}\ll\Omega\ll \Omega_{\rm c3}$ can be computed exactly to subleading order \cite{CPRY3}:

\begin{teo}[Energy between $\Omega_{c2}$ and $\Omega_{c3}$]
{\it If $\eps^{-1}\lesssim\Omega\ll\eps^{-4}$ as $\eps\to 0$, then}
$$E^{\rm GP}=E^{\rm TF}+\hbox{$\frac16$}\Omega |\log(\eps^4\Omega)|(1+o(1)).$$
\end{teo}
Here $E^{\rm TF}$ denotes the energy without the kinetic term. Below $\Omega_{2c}$ a similar formula holds (with a different scaling, $\eps^{-2/(s+2)}$ as length unit):

\begin{teo}[Energy between $\Omega_{c1}$ and $\Omega_{c2}$]
 If $|\log\eps|\ll\Omega'\lesssim\eps^{-1}$ as $\eps\to 0$, then
$$ E^{ \rm GP '}\label{energyprime}= E^{ \rm TF '}+\half \Omega' |\log(\eps^2{\Omega'})|(1+o(1)).$$
\end{teo}


\subsection{Vortices reduce kinetic energy}

The {potential term} $\sim\Omega^2$ {and the interaction term} $\sim\eps^{-2}$ {become comparable} when $$\Omega\sim\eps^{-1}.$$ This is the order of the  second critical speed $\Omega_{\mathrm c2}$ above which the {centrifugal force creates a \lq hole'}. 
In the sequel we shall focus on rotation speeds around and above $\Omega_{\mathrm c2}$ which means that $\Omega\gtrsim \eps^{-1}$. 
In this regime the kinetic energy term $\half |(\mathrm i\nabla +{\mathbf A})\Psi|^2$ is formally also of order $1/\eps^2$ if 
  $\Omega\sim 1/\eps$. {}%
Its contribution to the energy is, however, of { lower order}, namely $\sim\Omega|\log\eps|$, because { a lattice of vortices  emerges} as $\eps\to 0$ and reduces the kinetic energy as remarked in 2.4. {}%
  


\subsection{The giant vortex regime}

Consider a variational ansatz for the wave function of the form\footnote{For simplicity of notation we assume that $\Omega$ is an integer which is justified since $\Omega\to \infty$.} 
\begin{equation*}\label{ansatz}
 \psi(\xv)= g(\xv)\exp(\mathrm i\Omega\vartheta)\end{equation*}
with a {real valued} function $g$, normalized such that $\int g^2=1$.This  gives

\begin{equation*}\mathcal E^{\rm GP}[\psi]=\int_{\mathbb R^2}\left\{\half|\nabla g|^2+\half\Omega^2(x-x^{-1})^2g^2 +\Omega^2\left(\hbox{$\frac 1s$}x^s-\half x^2\right)g^2+\eps^{-2}g^4\right\}
\equiv{\mathcal E}^{\mathrm{ gv}}[g].
\end{equation*}
The unique positive minimizer $g_{\rm gv}$ of ${\mathcal E}^{\mathrm{ gv}}$ is  rotationally symmetric and we denote the corresponding energy by $E^{\rm gv}$. 

The following results are proved in \cite{CPRY2, CPRY3}.


\begin{teo}[Energy in the giant vortex regime] There is a constant  $0<\bar\Omega_0<\infty$ such that for $\Omega=\Omega_0\,\eps^{-4}$ with $\Omega_0>\bar\Omega_0$ the ground state energy is
 $$E^{\rm GP}\label{gvenergyas}=E^{\rm gv}+O(|\log\eps|^{9/2}).$$
 \end{teo}

\begin{teo}[Absence of vortices in the bulk]
{There is a constant $c>0$ such that for 
$\Omega=\Omega_0\,\eps^{-4}$ with $\Omega_0>\bar\Omega_0$ and $\eps$ sufficiently small 
the minimizer 
$\psi^{\rm GP}$ is {free of zeros in the annulus}} 
$$\mathcal A= \{\mathbf x:\, |1-x|\leq c\Omega^{-1/2}|\log\eps|^{1/2}\}.$$
\end{teo}


\subsection{On the proof of the GV transition}
The main issue is a precise lower bound to the energy. We restrict $\mathcal E^{\rm gv}$ to the annulus $\mathcal A$, obtaining a positive minimizer $g_0$. Define $u(\xv)$ on the annulus by writing
$$\psi^{\rm GP}(\mathbf x)=g_0 (x)u(\mathbf x)\exp(\mathrm i\Omega\vartheta).$$
The function $u$ contains all possible zeros of $\psi^{\rm GP}$ in the annulus.

 The variational equation for $g$ 
leads to the lower bound
\begin{equation*} E^{\rm GP}\geq E^{\rm gv}_{\mathcal A}+\mathcal E_{\mathcal A}[u]\end{equation*}
with a  functional of Ginzburg-Landau type with $g_0^2$ as weight: 
\begin{equation*} \mathcal E_{\mathcal A}[u]=\int_{\mathcal A}g_0^2\left\{\half|\nabla u|^2-\mathbf B\cdot \mathbf J(u)+\eps^{-2}g_0^2(1-|u|^2)^2\right\}\end{equation*}
where $\mathbf B=\Omega\,(x-x^{-1})\,\mathbf e_\vartheta$ and $\mathbf J(u)=\frac {\rm i}2(u\nabla u^*-u^*\nabla u)$.
The main task is to {estimate the negative term}  $-\int g_0^2\, \mathbf B\cdot \mathbf J(u)$.


For this purpose one writes
$g^2 \mathbf B=\nabla^\perp F$ with $\nabla^\perp=(-\partial_{x_2},\partial_{x_1})$ and a {potential function} $F$.  Integration by parts and  estimates of $F$ (this is the key point!) give
$$ \int_{\mathcal A}g^2\left\{\half|\nabla u|^2-\mathbf B\cdot \mathbf J(u)\right\} \geq
 -C\Omega_0^2|\log\eps|^{3/2}
$$
leading to the lower energy bound.
 
A consequence of this bound, combined with the variational upper bound $E_{\mathcal A}^{\rm gv}\leq 0$ is an {upper bound on the interaction term} for large $\Omega_0$: 
$$ \label{intbound}\int_{\mathcal A}\eps^{-2}g_0^4(1-|u|^2)^2\leq C\Omega_0^2|\log\eps|^{3/2}.$$

Together with the upper bound to the kinetic energy and standard inequalities this implies that $u$ must be close to 1, in particular free of zeros.


\subsection{Comparison with the \lq flat' case}

The flat case, $s=\infty$,  that is treated in detail in  \cite{CRY, CPRY1}, {differs from the case} $s<\infty$ in several respects:
\begin{itemize}
\item The GV transition takes place at $\Omega\sim \eps^{-2}|\log\eps|^{-1}$ rather than $\Omega\sim\eps^{-4}$.
\item The density profile in the GV regime is  of `Thomas-Fermi' type in the `flat' case, but for $s<\infty$ it is {gaussian} around $x=1$.
\item The \lq last' vortices before the GV transition have size $\sim \eps^{3/2}$ that is much smaller than the thickness of the annulus $\sim \eps|\log\eps|$.  For $s<\infty$ the size of vortices, $\sim\eps^2$ and the size of the annulus, $\sim\eps^2|\log\eps|^{1/2}$,  are almost comparable.
\end{itemize}{}
The techniques of proof in the two cases are also by necessity different: While {\it vortex ball constructions} and subsequent {\sl  jacobian estimates} (see \cite{SS}) for the potential function are applicable for the `small' vortices in a `flat' trap they are useless for $s<\infty$ and new ideas are required.


\subsection{Circulation and symmetry breaking}

At low rotation speeds below the onset of the second vortex the GP minimizer has rotationally symmetric density,  but a vortex lattice clearly breaks the symmetry. On the other hand, the giant vortex variational ansatz, that gives an excellent approximation to the energy and circulation for $\Omega_0>\bar \Omega_0$, is an eigenfunction of angular momentum. A true minimizer does not have this property, however:{}

\begin{teo} [Circulation and rotational symmetry breaking] {In the giant vortex regime $\Omega_0>\bar \Omega_0$  the circulation of any GP minimizer is  $2\pi\,\Omega+O(\Omega_0\,|\log\eps|^{9/4})$, but {no minimizer
is an eigenfunction of angular momentum}.}\end{teo}

Theis result holds both for $s<\infty$ and  $s=\infty$ \cite{CRY}--\cite{CPRY3}.



\section{Summary}

The study of the GP equation for dilute Bose gases in rotating, anharmonic  reveals a surprising rich landscape, both from the mathematical and physical point of view. Detailed analysis can be carried out in an asymptotic regime where both the coupling constant and the rotational speed are large.
Among the results found are:
\begin{itemize}
\item Energy asymptotics corresponding to a distribution of vorticity in a lattice of vortices for $\Omega_{c1}\ll\Omega\ll\Omega_{c3}$.
\item Emergence of a `hole' with strongly depleted density above a critical rotation speed $\Omega_{c2}\sim\eps^{-1}$.
\item Transition to a `giant vortex' state above $\Omega_{c3}\sim\eps^{-4}$ where the vortex lattice disappears from the bulk and all vorticity resides in the `hole', creating a macroscopic circulation in the bulk.
\item Breaking of rotational symmetry, also in the giant vortex regime.
\end{itemize}
\bigskip
\noindent
\small{{\bf Acknowledgements}

M.C. gratefully acknowledges financial support from the European Research Council under the European
UnionÕs Seventh Framework Programme ERC Starting Grant CoMBoS (grant agreement n. 239694).
N.R. was partially supported by the same programme (grant agreement MNIQS-258023).}


\end{document}